# Ion wake induced mode coupling in a horizontal chain in complex plasmas

Ke Qiao, Zhiyue Ding, Jorge R. Carmona, Jie Kong, Lorin S. Matthews and Truell W. Hyde

Center for Astrophysics, Space Physics and Engineering Research, Baylor University, Waco, Texas 76798-7283, USA

Ion wake induced mode coupling is investigated experimentally for a horizontal dust chain formed in a complex plasma, verifying results from previous simulation. A double branch of faint spectral lines is detected in the mode spectra which verifies the predicted rule of mode coupling between the vertical $z(j = i \pm 1)$ modes and the longitudinal mode $x(i)$. Discreet instabilities are observed as the branches of x- and z-modes intersect each other. The mode spectra in the vicinity of the instabilities exhibit enhanced energy density at specific coupled x and z modes, serving as direct evidence that these instabilities are caused by resonance between the coupled modes. The instability-induced melting threshold was found to obey the Lindemann criterion through analysis of the instantaneous relative interparticle distance fluctuation (IDF). The relation between mode spectra and dispersion relations was further studied by multiplying the mode spectra with a transition matrix connecting the bases of normal mode eigenvectors and Fourier series in $k$ space. Typical dispersion relations corresponding to the longitudinal and out-of-plane transverse Dust Lattice Waves (DLWs) are obtained, which also exhibit characteristics unique to finite systems, including discrete bands and strong fluctuations in the energy density.

## I. Introduction

Chainlike structures are ubiquitous in nature, in systems such as polyatomic molecules or proteins, in which the dynamics of mode couplings plays a major role. Unfortunately, mode analysis for these systems is difficult due to their fast time scales (picoseconds) and short length scales (nanometers). Complex (dusty) plasma system have therefore received increased attention over the past three decades in large part due to the fact that the constituent micron-sized particles and their motion can be directly with a CCD camera. As a result, phenomena such as phase transitions [1], wave propagation [2] and vibrational modes [3]–[7], which are of fundamental interest to traditional solid state/condensed matter systems can be investigated at the kinetic (atomic) level. As such, the study of complex plasma dust chains provides the potential to model mode coupling as well as other functions in such microscopic complex systems on a macroscopic time and length scale.

Complex (dusty) plasma experiments on earth typically examine dust particles levitated within the plasma sheath formed above the lower electrode of a capacitively-coupled rf discharge plasma. The sheath electric field which supports the negatively charged grains against gravity also causes the positive ions to stream towards the lower electrode. This ion flow creates an ion focusing regions with positive space charge downstream of each dust grain, known as the ion wake field [8]–[11]. The ion wake field contributes to mode couplings in both large 2D plasma crystals [12]–[14] and circular dust clusters [3], [4].

Recent research has shown that in large crystals (> 500 dust particles) the in-plane (longitudinal) dust lattice waves (DLWs) and out-of-plane (vertical) DLWs are coupled to each other due to the ion wakes crystals [12]–[14]. This can be seen in the phonon spectra of the plasma crystal, where an enhancement of the intensity at energies characteristic of longitudinal DLWs can also be observed in the spectra of the vertical DLWs crystals [12]–[14] When the longitudinal and vertical DLW branches intersect each other, a resonant instability occurs and the resulting phonon spectra shows a dramatic energy density enhancement at the cross section of the two branches. This corresponds to a dramatic increase of particle motion in the center of the plasma crystal (where the particles are most dense), which can lead to melting of the crystal. Interestingly, this is not, in general, a collective phenomenon specific to large crystals: other studies have found ion flow-induced mode coupling exists in small circular dust clusters with only 3-11 dust particles [3], [4]. In these clusters, the mode spectra exhibit a coupling pattern between individual horizontal and vertical normal modes, obeying coupling rules that have been derived theoretically and verified employing numerical simulation [3]. These studies show that when two coupled modes have the same frequency, a resonant instability can occur, where the energy density of both modes increases dramatically. Therefore, coupling between longitudinal and vertical DLWs in large crystals is inherently related to coupling between individual modes in finite clusters.

The above hypothesis was recently verified using a numerical simulation for wake-induced mode coupling in a horizontal linear chain [5]. With the ion wake effect modeled by a positive point charge downstream of each dust grain, the mode coupling in linear chains was found to demonstrate a much simpler and more straightforward coupling rule than in the case of circular clusters. In this case, coupling occurs between modes having adjacent indices, i.e. between modes with $i$ and $j = i \pm 1$, and $j$ and $i = j \pm 1$, where $i$ and $j$ are indices for the longitudinal and vertical modes, respectively. Dispersion relations were obtained through multiplication of the mode spectra by a transition matrix. The resulting dispersion relations and enhanced energy density corresponding to resonant instabilities were found to resemble those observed in large crystals. However, these simulation results were not verified experimentally, partly due to the difficulty of forming a horizontal linear chain with confinement strong enough for the coupling pattern to be detectable in the mode spectra.

In this research, horizontally aligned chains are formed in a glass box with dimensions 2.5 × 5 × 0.6 cm (height × length × width) placed on the lower electrode in a modified GEC reference cell. The box provides a confinement that is strong enough to enhance the inter-particle interaction allowing detection of the mode-coupling pattern. The mode spectra are obtained using thermally excited fluctuation of the particles, and patterns related to the mode coupling are identified. The coupling rules, resonance instabilities and resultant melting, and the relationship between the mode spectra and dispersion relations will all be examined.

## II. Method

The experiment reported here was carried out in a modified Gaseous Electronics Conference

(GEC) rf reference cell [15], which contains two 8-cm-diameter electrodes separated by a distance of 1.9 cm. The lower electrode is powered at 13.56 MHz while the upper ring-shaped electrode and chamber are grounded. A 2.5 ×5 ×0.6 cm (height ×length ×width) glass box was placed on the lower electrode to create the confinement potential necessary to establish the dust chain (Fig. 1 a). All experiments were conducted in an Argon plasma at 5.3 Pa employing rf powers between 5 and 30 W. Melamine formaldehyde (MF) particles (mass density of 1.51 g/cm$^3$ and diameter of 8.89 ± 0.09 μm) were used.

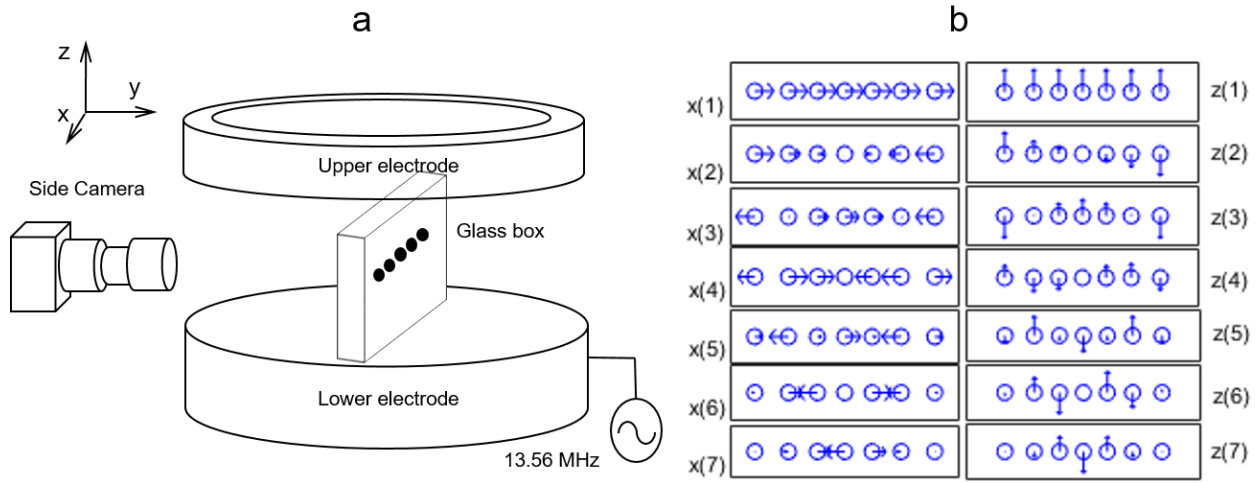

Fig. 1. (a) Lab setup (b) Eigenvectors for the *x* and *z* modes for a seven-particle chain.

Under these conditions, a balance between the gravitational force, interparticle force, and electric field forces created by the lower electrode and the walls of the box allowed formation of a single chain of dust particles aligned in the horizontal direction consisting of 6-8 MF dust particles within the glass box (Fig. 1 a). Particles were illuminated using a vertically fanned laser sheet. Once the dust chains were stabilized, the thermal fluctuation of the particles was recorded using a charge-coupled device (CCD) camera running at 60 frames per second and analyzed employing particle tracking [16] to obtain each particle's position and velocity for each frame. The normal mode spectra were then obtained by projecting the time series of the particle velocities onto the direction of the normal mode vectors for each pure mode (i.e., the eigenmodes calculated assuming a Yukawa interaction) and applying a Fourier transformation [3]–[5], [17].

The 1D geometry of the system allows the 3N (where N is the particle number) pure normal modes to be equally divided into x, y and z modes, with the eigenvectors directed along the x-, y- and z-axes [5]. Only the x and z modes were examined in this study because a recording of the top view motion synchronized with the side view motion cannot be achieved under current experimental conditions. The eigenvectors resemble a series of non-uniform standing waves with unequal wavelengths (Fig. 1 b); the N modes in each direction are indexed following the practice established in [5], *i.e.* in ascending order of the number of wavelengths. The modes in the *x* or *z* direction are denoted by x(*i*) or z(*j*), referring to the $i(j)^{th}$ x (z) mode.

## III. Mode coupling

Fig. 2 shows the longitudinal (x) and vertical (z) normal mode spectra for a seven- and eight-particle chain held at a rf power of 28 W, with the color (or gray scale) representing the logarithm of the power spectral density. Mode numbers 1~N correspond to the x(*i*) modes with *i* in ascending order from 1 to N, while mode numbers N+1~2N correspond to the z(*j*) modes with *j* in descending order from N to 1. Coupling between the x and z modes can be identified by detecting x and z modes having equal spectral frequencies [3]–[5]. As can be seen in the mode spectra of both the seven and eight particle chains, most frequencies corresponding to mode x(*i*) have two dim spectral lines corresponding to modes z(*j*) with $j = i \pm 1$. As a consequence, a double branch of dim spectral lines appear in the mode spectra below the z mode branch, with the higher branch corresponding to modes z($j = i - 1$) and the lower branch corresponding to modes z($j = i + 1$). These two branches, as well as the rules of coupling $j = i \pm 1$, show excellent agreement with previous predictions from numerical simulations [5]. Due to the faintness of the spectral lines and large signal-to-noise ratio, the corresponding coupling rule of $i = j \pm 1$ cannot be definitively verified.

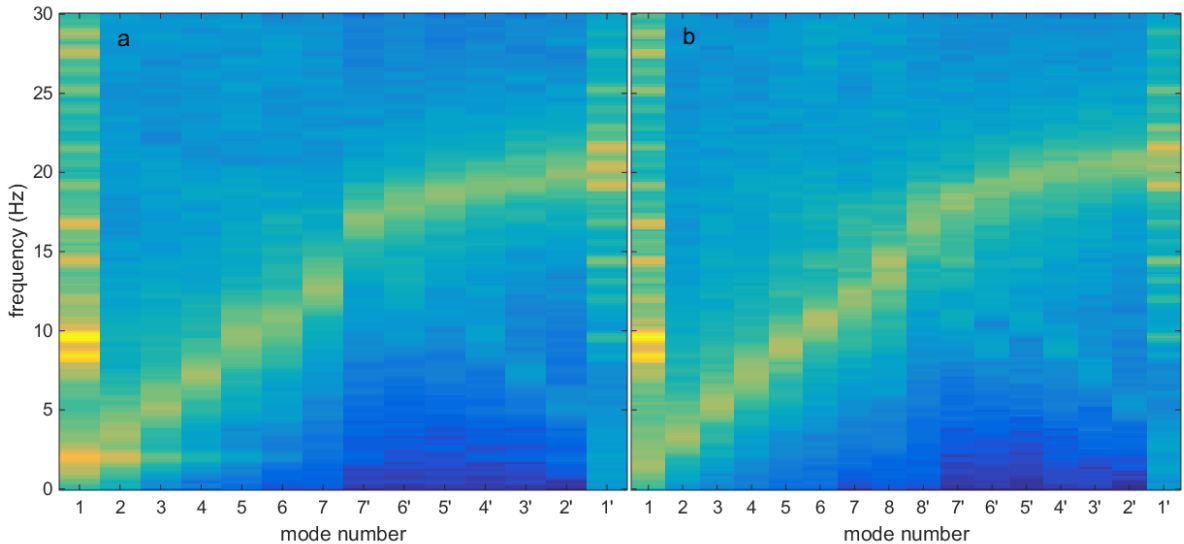

Fig. 2. (Color online) Normal mode spectra for (a) seven and (b) eight-particle chains at a rf power of 28 W. The first set of mode numbers correspond to the longitudinal *x*-modes, while the primes denote the vertical *z*-modes.

## IV. Discrete resonance instabilities and melting

The coupling of the $x(i)$ mode with modes $z(j = i \pm 1)$ implies that the $x(i)$ mode is no longer a pure horizontal mode, but rather a mode with mixed polarization, consisting of both the horizontal $x(i)$, and vertical $z(j = i + 1)$ and $z(j = i - 1)$ components. Therefore, resonant instabilities will occur when the frequencies of $x(i)$ and $z(j = i + 1)$ (or $x(i)$ and $z(j = i + 1)$) approach each other. If the resulting instability is strong enough, it can lead to melting of the chain structure [5]. To verify this prediction, we examined the sloshing mode frequencies on the $x$- and $z$-directions, which can be considered as a measurement of the confinement strength in the corresponding direction. The relative position between the $x$ and $z$ mode branches in the mode spectra can be tuned through variation of the rf power. As the rf power was varied from 18 W to 12 W, the vertical sloshing mode $z(1)$ frequency decreased monotonically while the horizontal sloshing mode $x(1)$ frequency barely changed [4], [8]. Therefore, as the rf power decreases from a high value, the $x$ and $z$ branches approach and intersect each other.

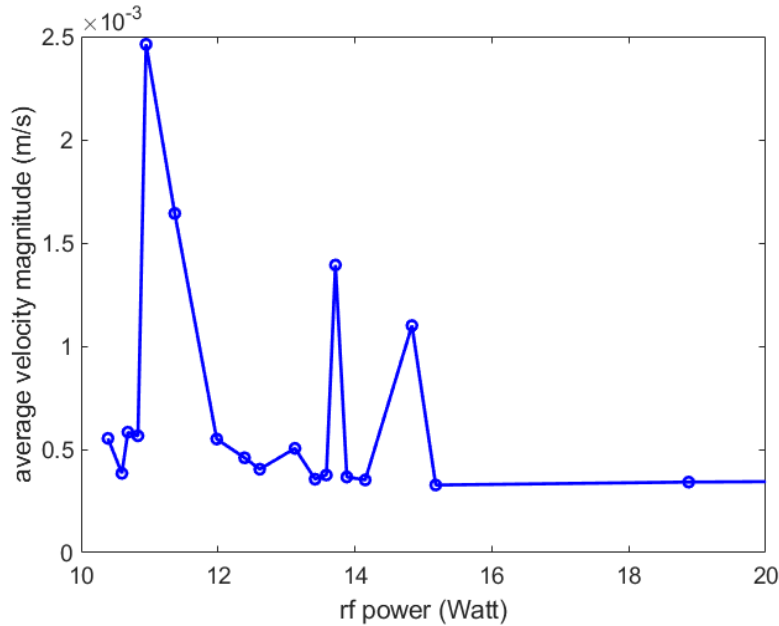

Fig. 3. Magnitude of the average particle velocity for an eight-particle chain as a function of the rf power.

Discreet instabilities were observed for chains of $N \geq 6$ as the rf power was decreased from an initial value of 20 W. Using an eight-particle chain as an example, it was observed that the chain structure remained stable for rf power $\geq 16$ W, exhibiting its first instability at approximately 15 W. Three discrete instabilities occured as the power decreased, as identified by both particle trajectories [see supplemental movie] and the magnitude of the average particle velocity (Fig. 3). The average velocity for the eight particles in Fig. 3 clearly shows three peaks (maxima) as the rf power decreases from 15 W to 10 W (corresponding to a change in the vertical sloshing mode frequency from 21 Hz ~ 16 Hz). Measurements were only made for rf power > 10.5 W because at lower power the dust system spreads in the vertical direction and the system structure loses its quasi-1D characteristic.

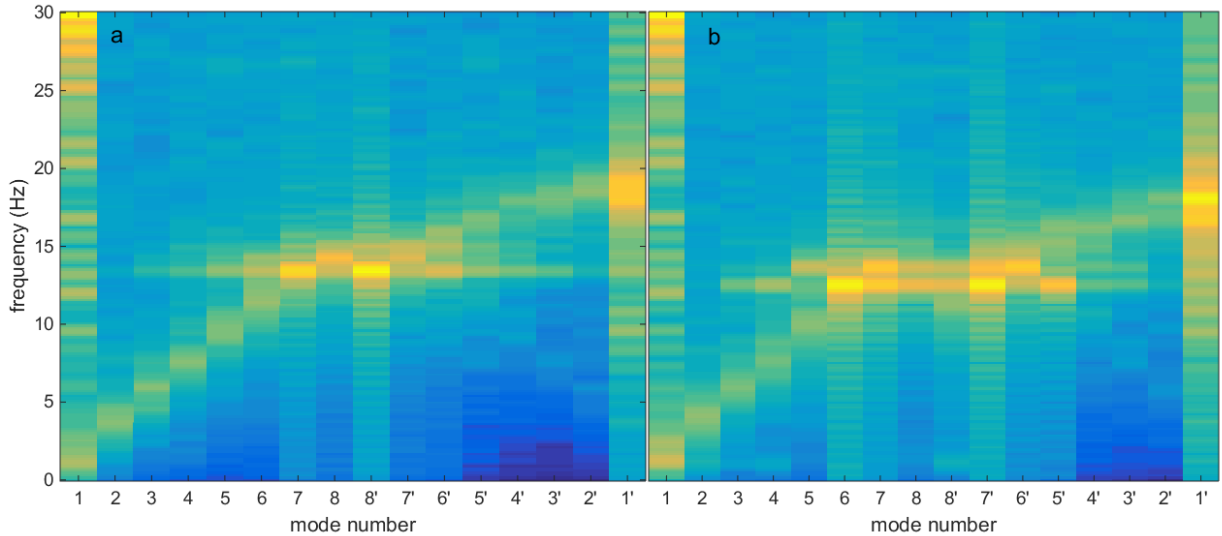

Fig. 4. (Color online) Normal mode spectra for an eight-particle chain at the first and second resonance instabilities with power equal to (a) 14.5 W and (b) 13.8 W.

The mode spectra obtained at powers associated with the first two peaks, 14.5 W and 13.8 W respectively, are shown in Fig. 4. The mode spectrum corresponding to the first peak shows "hot spots" at modes $x(7)$ and $z(8)$ (Fig. 4 a). The energy density is also enhanced at modes $x(8)$ and $z(7)$, but to a lesser degree, indicating that the instability is caused primarily by the resonance between modes $x(7)$ and $z(8)$ with the resonance between modes $x(8)$ and $z(7)$ providing a minor contribution. Correspondingly, a clear horizontal cascade is seen at the frequency of modes $x(7)$ and $z(8)$ spread over all mode numbers, while a much fainter cascade can be seen at the frequency of modes $x(8)$~$z(7)$. This phenomenon of simultaneous excitation of two different but neighboring resonances was identified in previous numerical simulations as a resonance overlap [5], but has only been observed in chains having much larger particle numbers (for example, $N = 20$).

At the power setting for the second instability (Fig 4 b), the frequencies associated with the z mode branch decrease further and the "hot spots" in the spectrum become modes $x(6)$ and $z(7)$, while modes $x(7)$ and $z(6)$ are also enhanced. This indicates that the second instability is mainly due to the $x(6)$~$z(7)$ resonance, although the $x(7)$~$z(6)$ resonance also contributes. Again, the $x(6)$~$z(7)$ resonance produces a clear cascade in the mode spectra while the $x(7)$~$z(6)$ resonance causes a fainter one. These cascades are generally recognized to be caused by nonlinear effects and/or chain resonances between multiple coupled x- and z-modes[3]–[5], [12]–[14].

As mentioned above, the simultaneous excitation of two resonances (i.e. resonance overlap) occurs experimentally at much lower particle numbers than appeared in the numerical simulation

[5], where resonances are much more clear-cut and distinguishable. This is apparently due to the non-ideal conditions seen in real experiments, such as the variation in particle size and system fluctuations. Further reduction of the rf power causes the resonance overlap to rapidly become dominant. For this reason, only the first three instabilities (caused by overlapped resonances) can be distinguished.

The melting caused by the first resonant instability can be clearly observed. The melting threshold can be identified by the transition from a state where all particles vibrate around their equilibrium positions to a state where particles are "hopping" between their equilibrium positions (see the supplemental movie, time t = 1:20). This microscopic view has been employed in extensive research on finite system melting [18]–[21].

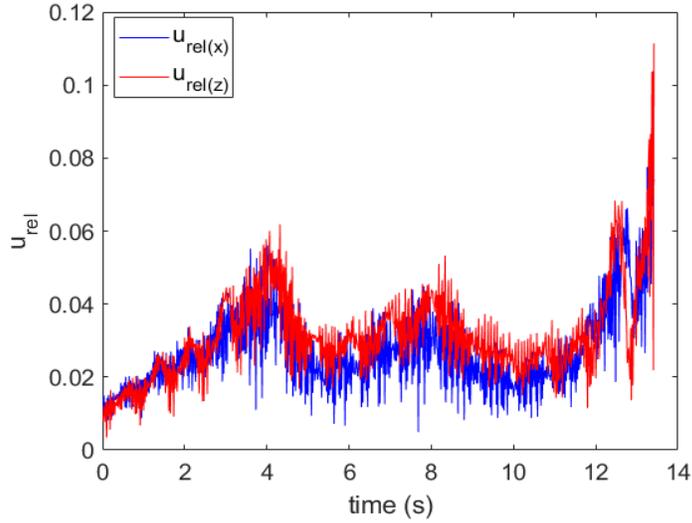

Fig. 5. (Color online) $u_{rel}(\sigma)$ as a function of time for the melting of an eight-particle chain induced by the first resonance instability.

Quantitatively, the temporal evolution of the melting process can also be investigated employing the instantaneous relative interparticle distance fluctuation (IDF) $u_{rel}$ [3], [22] as defined by

$$u_{rel(\sigma)} = \frac{2}{N(N-1)} \sum_{1 \leq i \leq j}^{N} \frac{|\sigma_{ij} - \sigma_{ij0}|}{r_{ij0}}, \quad (1)$$

where $\sigma$ can be either x or z; $N$ is the particle number; $\sigma_{i,j}$ and $\sigma_{i,j0}$ are the instantaneous and equilibrium distances between particles $i$ and $j$ along the specified direction; and $r_{ij0} = \sqrt{x_{ij0}^2 + y_{ij0}^2 + z_{ij0}^2}$. As a representative example, Fig. 5 shows $u_{rel(\sigma)}$ as a function of time for the melting of an eight-particle chain at the first instability. The function $u_{rel(\sigma)}$ is plotted up to the point in time when the melting threshold is reached, where the particles begin to hop between equilibrium positions. As can be seen, both $u_{rel(x)}$ and $u_{rel(z)}$ exhibit the same trends with time, as predicted by numerical simulations [5]. The value of $u_{rel}$ at the melting threshold is close to 0.1 for motion in both x and z, in excellent agreement with the Lindemann criterion [22]. This experimentally verifies the prediction based on the simulation results that the Lindemann criterion, which was originally used for equilibrium melting driven by an increase in temperature, is also suitable for nonequilibrium melting driven by resonance instabilities, even for extremely anisotropic systems such as dust chains within a complex plasma.

## V. Relationship between Mode Spectra and Dispersion Relations

The numerical simulations in [5] also showed that the dispersion relation in a large (infinite) system is inherently related to the mode spectra of a finite system. For this case, the dispersion relations for DLWs were obtained by multiplying the mode spectra matrix with a transition matrix defined as

$$U_{n,k(\sigma)} = \sum_{i=1}^{N} e_{i,n(\sigma)} e^{-ikx_i}, \qquad (2)$$

where $e_{i,n(\sigma)}$ is the eigenvector for the $n^{\text{th}}$ $\sigma$ mode at particle position $i$, and $k$ is a specified wave number. Thus, the transition matrix $U_{n,k}$ represents the transformation between two bases, the base of normal mode eigenvectors and the Fourier series in $k$ space.

We applied this technique to our experimental results from this study to obtain the dispersion relations. Fig. 6 shows the patterns obtained from the transition of the mode spectra produced by an eight-particle chain at the second resonance instability. As shown in Figs. 6 (a) and (b), the patterns obtained from the and z modes resemble typical dispersion relations corresponding to the longitudinal and out-of-plane transverse DLWs, respectively. The longitudinal DLW shows an acoustic characteristics (frequency increases as $k$ increases) for low values of $k$ while the out-of-plane DLW shows an optical characteristics (frequency decreases as $k$ increases), with both DLWs exhibiting rising and falling branches. A clear signature of the mode coupling is evident in the trace of the longitudinal dispersion seen in the pattern obtained from the z modes (Fig. 6 b). The energy density is enhanced dramatically at the intersection of the two branches, clearly resembling the hybrid wave modes observed previously only in large crystals [12]–[14].

On the other hand, the dispersion relations also exhibit characteristics caused by the finite nature of the system, as predicted in [5].

1) Branches are formed in discrete stripes within the frequency regime. This is a well-known characteristic of a finite structure since the stripes correspond to the discrete normal modes.
2) The energy density exhibits strong fluctuations between the rising and falling branches. As discussed in [5], fluctuations in energy density can be most easily explained by the transition matrix $U_{n,k}$. As illustrated in Fig. 7, the transition matrix in this case exhibits strong intensity fluctuations over the $k$ regime for each mode number. The fluctuation (series of maxima) is caused by the non-uniform wave characteristic of the normal modes (Fig. 1 b), which are in turn induced by the finite nature of the system. This argument is further supported by a simulation conducted in [5], where the fluctuation is clearly reduced when the central section of a long chain is investigated.

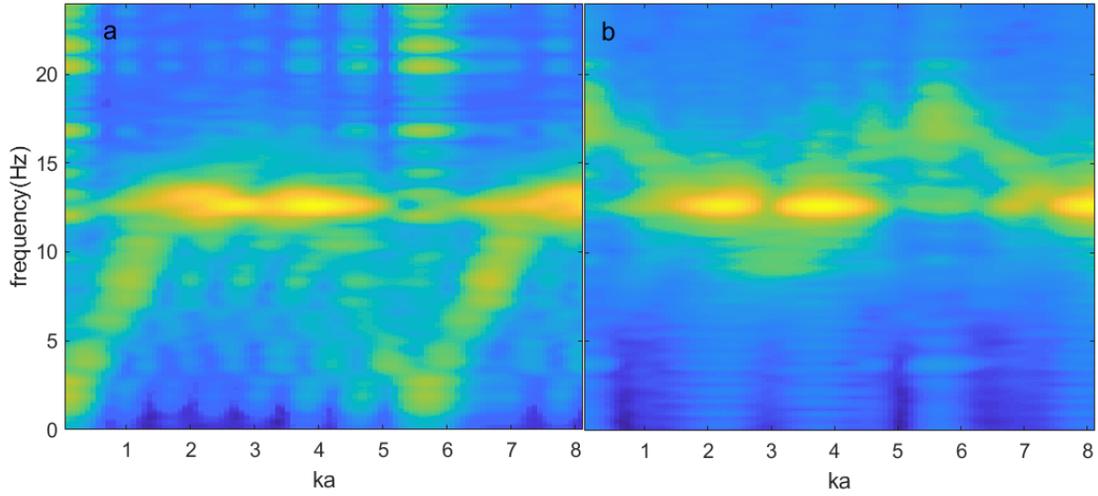

Fig. 6. Dispersion relations obtained from the (a) *x* and (b) *z* mode spectra.

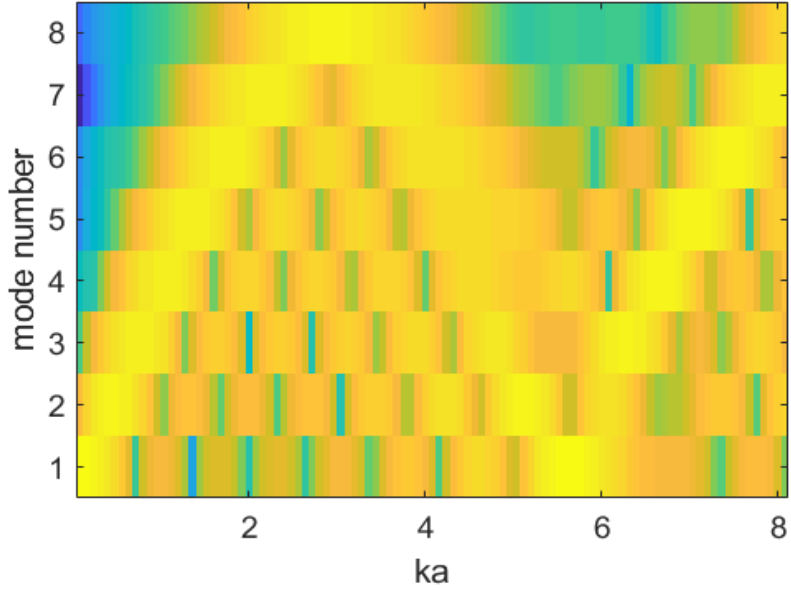

Fig. 7. The transition matrix for an eight-particle chain.

## VI. Summary

This research experimentally verified the ion wake-induced mode coupling in horizontal chains predicted by numerical simulations [5]. The horizontal chain structures were realized experimentally by using a long, narrow glass box placed on the lower electrode to provide strong horizontal confinement, allowing the inter-particle interaction to be strong enough for the mode coupling to be detected. A double branch of faint spectral lines found in the mode spectra verifies the rule of mode coupling of $z(j = i \pm 1)$ with $x(i)$. By tuning the rf power from high to low (decreasing the vertical confinement), the branches of the x and z modes in the power spectra can be seen to approach and intersect each other. Consequently, discreet instabilities occur, as identified from particle trajectories with dramatic increases in the average particle velocity. The mode spectra in the vicinity of the instabilities exhibit "hot spots" (enhanced energy density) at specific coupled $x$ and $z$ modes which have the same frequency, serving as direct evidence that these instabilities are caused by resonance between the coupled modes.

The melting process caused by these resonance instabilities was investigated employing the instantaneous relative interparticle distance fluctuation (IDF), and the melting threshold was found to be in good agreement with the Lindemann criterion. Finally, the relationship between the mode spectra and the dispersion relations was studied by multiplying the mode spectra with a transition matrix connecting the bases of normal mode eigenvectors and Fourier series in $k$ space. The pattern obtained resembles typical dispersion relations corresponding to the longitudinal and out-of-plane transverse DLWs. It also exhibits characteristics unique to finite systems, including DLW branches formed in discrete stripes and strong fluctuations in the energy density.

The experimental results provide good agreement with predicted results obtained from simulation [5]. In the experimental results, the resonances are less distinguishable and begin to overlap for fewer particles than in the simulation. As a result, only the first three resonance instabilities can be identified. For the same reason, the coupling rule x($i = j \pm 1$) with z($j$) cannot be definitely identified, although a faint pattern corresponding to these branches was observed in the mode spectra. The differences observed are primarily due to the non-ideal conditions inherent in experiments, such as the variance in particle size and system fluctuations. Another contributing factor is the nonlinear interaction between the ion wakes of adjacent particles, which can vary considerably as particles change their relative positions [23]. At the present time, the coupling of particle oscillations in the $xz$-plane to oscillations in the $y$-direction ($y$-mode) cannot be studied because the current experimental setup is not capable of producing a synchronized recording of the side- and top-views. Despite these experimental limits, the results on ion wake-induced mode coupling and related phenomena shown here should shed light on functions in other chainlike microscopic complex systems such as polyatomic molecules and proteins. If so, this will allow for research in areas such as intramolecular vibrational redistribution (IVR) in polyatomic molecules [24] and the internal resonance in nonlinear dynamical systems [6], [25] using complex plasmas as an analog.

Acknowledgements
Support from NSF Grant No. 1707215, No. 1740203 and NASA/JPL 1571701 is gratefully acknowledged.